\documentclass[aip, apl,amsmath,amssymb,reprint,]{revtex4-1}
\usepackage{graphicx}

\begin{document}

\title{Local photocurrent generation in thin films of the topological insulator Bi$_{2}$Se$_{3}$}

\author{C. Kastl}
\affiliation{Walter Schottky Institut and Physik-Department, Technische Universit\"at M\"unchen, Am Coulombwall 4a, 85748 Garching, Germany.}
\author{T. Guan}
\author{X. Y. He}
\author{K. H. Wu}
\author{Y. Q. Li}
\affiliation{Institute of Physics, Chinese Academy of Sciences, Beijing 100190, China.}
\author{A. W. Holleitner}
\affiliation{Walter Schottky Institut and Physik-Department, Technische Universit\"at M\"unchen, Am Coulombwall 4a, 85748 Garching, Germany.}

\date{\today}

\begin{abstract}
We report on the optoelectronic properties of thin films of Bi$_2$Se$_3$ grown by molecular beam epitaxy. The films are patterned into circuits with typical extensions of tens of microns. In spatially resolved experiments, we observe submicron photocurrent patterns with positive and negative amplitude. The patterns are independent of the applied bias voltage, but they depend on the width of the circuits. We interpret the patterns to originate from a local photocurrent generation due to potential fluctuations.
\end{abstract}

\maketitle

In recent years, a new class of solid state materials, called topological insulators, has emerged. \cite{kane2005} In the bulk, a topological insulator behaves like an ordinary insulator exhibiting states with a band gap between valence and conduction band. At the surface, gapless states exist showing remarkable properties such as helical Dirac dispersion near zero energy \cite{hsieh2009} and suppression of backscattering. \cite{zhang2009,roushan2009} Initially, the existence of such states was predicted theoretically \cite{bernevig2006} and verified experimentally \cite{konig2007} for two-dimensional HgTe/Hg quantum wells. Later on, also three-dimensional topological insulators, \cite{fu2007,Hsieh2008,chen2009,zhang2009} such as Bi$_2$Se$_3$, \cite{zhang2010a,xia2009} were found. In the three-dimensional systems, the characterization of the surface states via transport experiments is often hindered by a large residual bulk charge carrier density. \cite{taskin2009,checkelsky2009,analytis2010} However, both theoretical \cite{hosur2011,dora2012}  and experimental work \cite{mciver2011} provides evidence that the helical surface states can be selectively addressed by excitation with circularly polarized light and subsequently read out in a corresponding photocurrent measurement.

Here, we experimentally investigate photocurrents generated in thin films of Bi$_2$Se$_3$ grown by molecular beam epitaxy. By measuring the photocurrent at zero bias voltage, we find a global photoresponse of the overall circuit, whose average magnitude depends only on the device geometry and excitation intensity. Remarkably, the photocurrent shows reproducible submicron fluctuations with positive and negative amplitude. By changing the charge carrier density via a back gate, these spatial photocurrent patterns can be continuously deformed. Therefore, we conclude that the photocurrent is generated by local fluctuations of the potential landscape in  Bi$_2$Se$_3$. Our findings are consistent with a photoresponse predicted very recently for gapless materials, \cite{song2011} and they may prove useful for the design of photodetectors based on topological insulators.\cite{mciver2011} Furthermore, such information on fluctuations may be valuable for understanding electron transport properties such as linear magnetoresistance \cite{Zhang2011, He2012a, Gao2012, Wang2012a, He2012, Parish2003, Abrikosov1998,Wang2012b} as well as for the study of the exotic physics near the Dirac point in topological insulators. \cite{Martin2007}

Starting point are thin films of Bi$_2$Se$_3$ grown by molecular beam epitaxy on SrTiO$_3$(111). \cite{chen2010} The films are patterned into  50 $\mu$m wide Hall bar devices using standard optical lithography and plasma etching. Ohmic contacts and a gate electrode located at the backside of the substrate are formed by metallization of chromium and gold. We investigate samples from two different wafers with a film thickness of 10 nm and 20 nm, respectively, which yield qualitatively identical results. By independent magneto-transport measurements, devices fabricated from similar wafers were shown to exhibit indications of the surface transport characteristics in topological insulators. \cite{chen2010,chen2011}

All photocurrent measurements are carried out at temperatures around $T \approx 30-40$~K in a He atmosphere at a pressure of about 5 mbar using a scanning confocal laser microscope ($\lambda$ = 800~nm). With a spot diameter of ~1.5~$\mu$m and an integrated laser power of 10~$\mu$W, the light intensity $I_\text{opt}$ is on the order of kW/cm$^2$. The laser is modulated at a chopper frequency $f_\text{chop}$. The resulting photocurrent is measured between two electrically unbiased contacts using a current-voltage converter and a lock-in amplifier triggered at  $f_\text{chop}$. By scanning the laser spot across the sample and simultaneously detecting the photocurrent and the intensity of the reflected light, corresponding photocurrent and reflectance maps are recorded.

\begin{figure}
\includegraphics{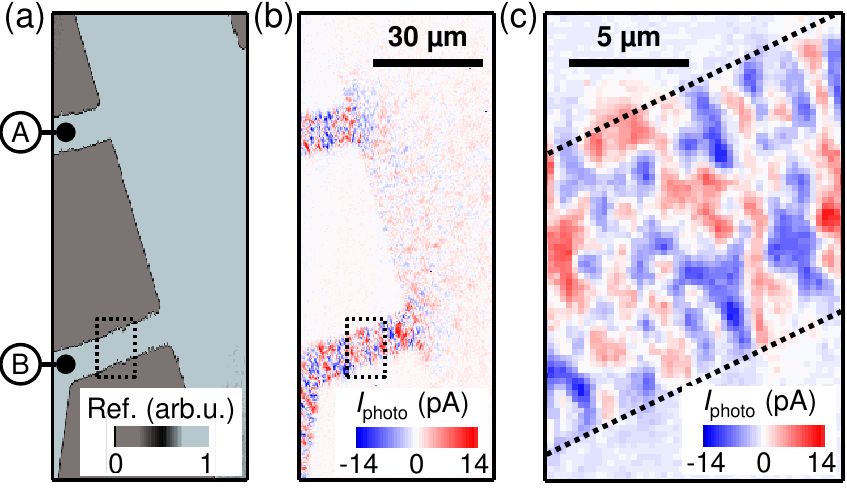}
\caption{\label{fig:01}(color online): (a) Reflectance map of a 10 nm thin film of Bi$_2$Se$_3$ (bright) on a SrTiO$_3$ substrate (dark). Photocurrent $I_\text{photo}$ is measured between two unbiased contacts A and B as a function of excitation position. (b) Photocurrent map of same area as in (a) ($f_\text{chop}= 239$~Hz,  $I_\text{opt} = 1.2$~kW/cm$^2$). (c) Photocurrent map of the area marked by dotted rectangle. Dotted lines indicate boundary of channel  ($f_\text{chop}=239$~Hz,  $I_\text{opt} = 0.9$~kW/cm$^2$).}
\end{figure}

Fig.~\ref{fig:01}(a) depicts such a reflectance map of a circuit made out of the 10 nm thin film with unbiased contacts A and B. Contact A is kept at zero potential whereas contact B is connected to a current-voltage converter. All other contacts of the circuit are floating (not shown). The corresponding photocurrent map is depicted in Fig.~\ref{fig:01}(b). We find distinct photocurrent patterns with positive and negative photocurrent amplitudes on the order of $|I_\text{photo}| \approx 10$~pA at $I_\text{opt} = 1.2$~kW/cm$^2$. The signal is concentrated along a current path between the non-floating contacts A and B, otherwise it fades out towards an average amplitude around zero. Furthermore, the photoresponse is enhanced when the geometry of the Bi$_2$Se$_3$ channel is narrowed [Fig.~\ref{fig:01}(b)]. A magnified high resolution scan of the area marked by the dotted rectangle is depicted in Fig.~\ref{fig:01}(c). 

Relaxation and recombination times in Bi$_2$Se$_3$ are on the order of ps. \cite{Hsieh2011,kumar2011} Therefore, primary photoexcited carriers typically decay and thermalize back to equilibrium conditions before reaching the contacts. In our case, the contacts are located at a macroscopic distance of about 1~mm from the excitation spot. Therefore, we can exclude direct photocurrents as well as photothermoelectric currents, as seen near contact interfaces in Bi$_2$Se$_3$ \cite{mciver2011} and also graphene. \cite{lee2008,prechtel2012} Since the contacts A and B are unbiased, we can furthermore exclude persistent photoconductance effects to cause the observed photocurrent patterns. \cite{Hof2008} Instead, we attribute our findings to a global photoresponse caused by a local photocurrent generation mechanism as recently proposed for topological insulators. \cite{song2011} According to this model, a local, microscopic photocurrent $\boldsymbol{j}_\text{loc}(\boldsymbol{r})$ sets up a global photocurrent $I_\text{global}$ given by
\begin{equation}
\label{eq:01}
I_\text{global}=A\cdot \int d^2 r \,  \boldsymbol{j}_\text{loc} (\boldsymbol{r}) \nabla \psi (\boldsymbol{r})	
\end{equation}
where $A$ is a prefactor accounting for the resistance of sample and circuitry and  $\psi (\boldsymbol{r})$ is an appropriate weighting field. \cite{song2011} We observe that the photocurrent patterns are very reproducible and persist on samples from both wafers independent of  $f_\text{chop}$ (up to 4~kHz), temperature (up to room temperature) and applied bias voltage (up to 100~mV) (data not shown), and that $|I_\text{photo}|$ depends linearly on $I_\text{opt}$, which all is consistent with the above model.

\begin{figure}
\includegraphics{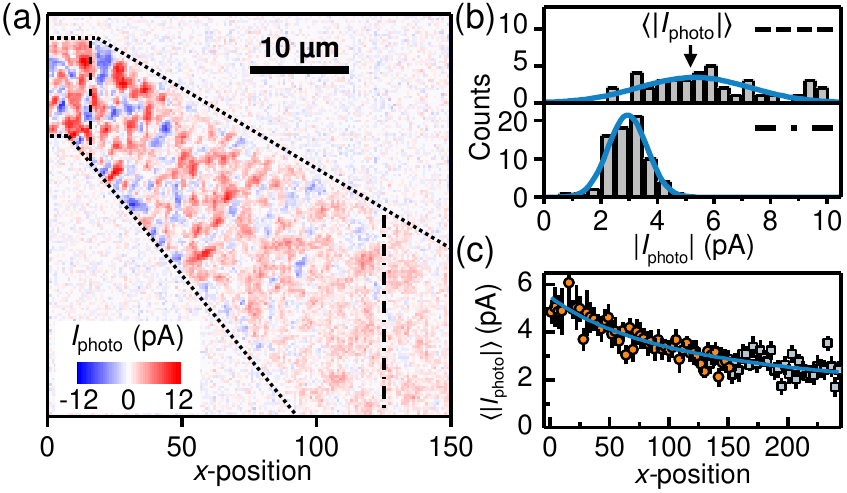}
\caption{\label{fig:02}(color online): Photocurrent map of a 10 nm thin Bi$_2$Se$_3$ channel with varying lateral width. Dotted lines indicate boundary of channel.  (b) Histograms of $| I_\text{photo} |$ exemplarily depicted along dashed and dashed-dotted lines in (a). (c) $\langle | I_\text{photo} | \rangle$ as a function of $x$-position. Squares correspond to a second photocurrent map shifted compared to (a) ($f_\text{chop}=439$~Hz,  $I_\text{opt} = 0.56$~kW/cm$^2$).}
\end{figure}

We systematically investigate the impact of device geometry on $I_\text{photo}$, which is already evident in Fig.~\ref{fig:01}(b), by varying the width of the channels. Fig.~\ref{fig:02}(a) shows a photocurrent scan along a narrowing Bi$_2$Se$_3$ channel. Clearly, the average magnitude of the photocurrent patterns decreases with increasing width. In particular, Fig.~\ref{fig:02}(b) depicts histograms of the photocurrent amplitude $|I_\text{photo}|$  inside the channel along the dashed and dashed-dotted line in Fig.~\ref{fig:02}(a). By approximating the peak with a Gaussian distribution, the mean amplitude $\langle | I_\text{photo} | \rangle$ of the patterns can be quantified. Fig.~\ref{fig:02}(c) shows $\langle | I_\text{photo} | \rangle$  as a function of the $x$-position along the channel. The error bars denote the standard deviation of the fitted distributions. With increasing width of the channel, the average magnitude of the photocurrent decreases. The solid line is a decrease curve with a saturation value at the finite noise level of 0.7 pA.

Generally, after the local photocurrent generation, i.e. the photogenerated charge carriers are spatially separated, a lateral diffusion of these charge carriers takes place until they finally relax back to equilibrium conditions. Assuming typical values of $D=10$~cm$^2$/s and $\tau_\text{relax}=10$~ps for the diffusion constant and the relaxation time, respectively, \cite{Hsieh2011,kumar2011,sobota2012,wang2012} one can estimate the effective diffusion length of the photogenerated charge carriers to be on the order of 300 nm. \cite{Smith1988} However, the photocurrent depends on device geometry even if the excitation position is separated by more than 10 $\mu$m from the channel boundary [Fig.~\ref{fig:01}(b), Fig.~\ref{fig:02}]. Therefore, we attribute the decrease of the photocurrent magnitude with increasing channel width to the sensitivity of the long-range global photocurrent $I_\text{global}$ to the boundary conditions, in accordance with Eq.~\ref{eq:01}.

\begin{figure}
\includegraphics{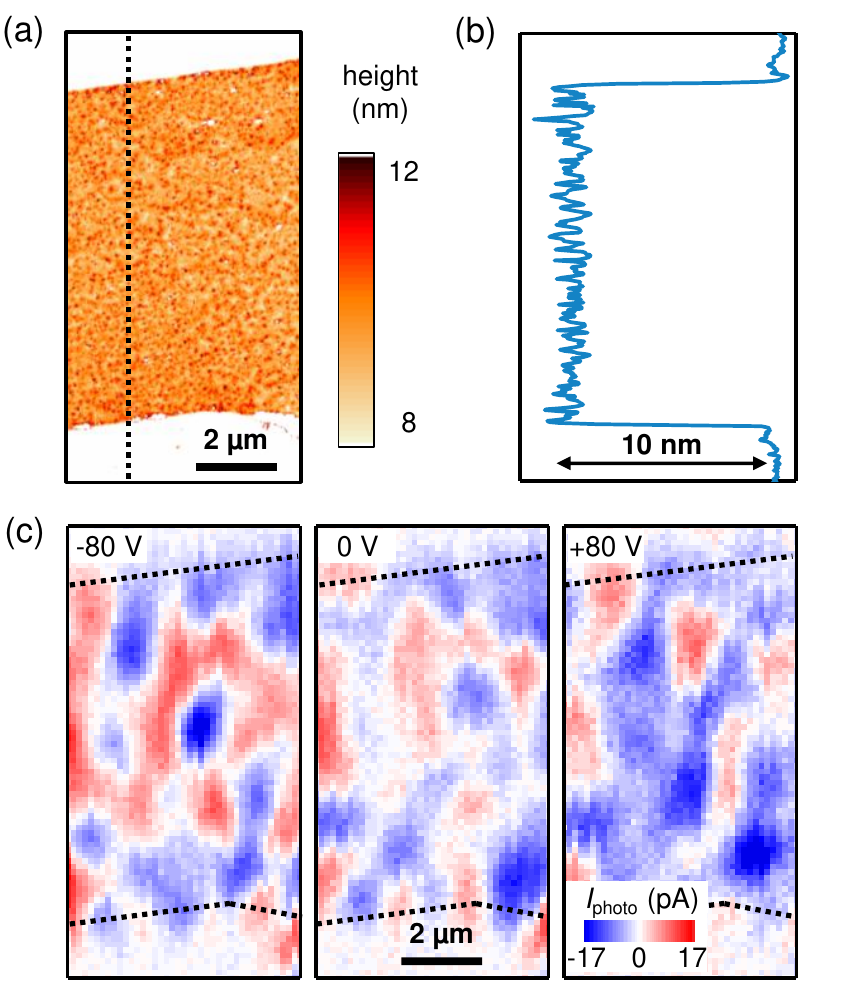}
\caption{\label{fig:03}(color online): (a) AFM image of a 10 nm thin Bi$_2$Se$_3$ film. (b) Single line scan along dotted line in (a). (c) Photocurrent maps of same area as in (a) for back gate voltages of -80~V, 0~V and +80~V, respectively. Dotted lines indicate sample boundary ($f_\text{chop}=439$~Hz,  $I_\text{opt} = 0.16$~kW/cm$^2$).}
\end{figure}
From atomic force microscopy (AFM), we find that the surfaces of the Bi$_2$Se$_3$ films are homogenous in height up to a single quintuple layer of Bi$_2$Se$_3$ [Figs.~\ref{fig:03}(a) and (b)]. Photocurrent maps of the same area are depicted in Fig.~\ref{fig:03}(c). One cannot relate fluctuations in photocurrent, especially changes in sign, to fluctuations in topography. However, gate dependent measurements reveal that the observed photocurrent patterns can be continuously deformed by changing the charge carrier density. For the gate voltages shown in Fig.~\ref{fig:03}(c), the charge carrier density is varied by approximately $\Delta n \approx 10^{13}$~cm$^{-2}$ compared to a bulk sheet density $n_0 \approx 2\cdot10^{13}$~cm$^{-2}$ in these devices. \cite{chen2010} The voltage-tunability suggests that the local photocurrent generation is related to the electrostatic potential landscape in the Bi$_2$Se$_3$ films. Scanning tunneling microscopy (STM) experiments on thin films of Bi$_2$Se$_3$ and Bi$_2$Te$_3$ revealed potential fluctuations on the order of tens of meV with a spatial extension of several nm. \cite{beidenkopf2011} Therefore, we attribute the local photocurrent patterns to be generated by the separation of photogenerated electrons and holes due to the fields arising from such potential fluctuations. The direction of the photocurrent and consequently the sign of the macroscopic current are then determined by the direction of the potential gradient. We note that the observed pattern size of the photocurrent is on the order of 1~$\mu$m and below as can be seen in Figs.~\ref{fig:01}(b,c), \ref{fig:02}(a) and \ref{fig:03}(c). In our interpretation, this size is determined by the interplay of the potential fluctuations, the laser spot profile, and the diffusion of the photogenerated charge carriers.
 
We note that the observation of a local photocurrent generation points towards a current incompressibility as predicted in gapless materials. \cite{song2011} Thin films of Bi$_2$Se$_3$ are usually n-type directly after growth due to selenium vacancies \cite{navratil2004} and they suffer from additional n-doping due to surface degradation when exposed to atmospheric conditions \cite{analytis2010,kong2011} resulting in a large residual bulk electron density. Even with electrostatic gating up to $V_\text{gate} = -80$~V, the chemical potential cannot be moved into the bulk band gap across the whole thickness of the film.\cite{chen2010,chen2011}  The participation of the bulk carriers in the transport may explain why the photocurrent patterns in Fig.~\ref{fig:03}(c) do not change qualitatively with the gate voltage. Clearer results concerning the surface states might be achieved using material systems where intrinsic doping effects are compensated,\cite{He2012} such that the bulk transport can be suppressed.

In summary, we report on spatially resolved photocurrent measurements of molecular beam epitaxy grown films of the topological insulator Bi$_2$Se$_3$. We verify a global photoresponse mechanism predicted for gapless materials caused by a local photocurrent generation. We interpret the local photocurrent to be generated by spatial fluctuations of the electrostatic potential.

We gratefully acknowledge support from the DFG via grant HO 3324/4 and and from National Basic Research Program of China (Project No. 2012CB921703), National Science Foundation of China (Project Nos. 91121003 and 10974240) and the Chinese Academy of Sciencese.

\end{document}